\documentclass{emulateapj}
\newcommand{\lsim}{\mbox{$_<\atop^{\sim}$}}

\newcommand{\mum}{$\,\mu$m}
\newcommand{\muJy}{$\,\mu$Jy}

\newcommand{\spitzer}{\textsl{Spitzer}}

\slugcomment{Accepted by ApJ}

\shorttitle{Interacting \& Merging Galaxies: Spitzer's View}
\shortauthors{Bridge et al.}

\begin{document}

\title{The Role of Galaxy Interactions and Mergers in Star Formation at $z\le1.3$: Mid-Infrared Properties in the \textsl{Spitzer} First Look Survey \altaffilmark{1}}

\author{
C.R.\ Bridge,$\!$\altaffilmark{2}
P.N.\ Appleton,$\!$\altaffilmark{3}
C.J.\ Conselice,$\!$\altaffilmark{4}
P.I\ Choi,$\!$\altaffilmark{5}
L.\ Armus,$\!$\altaffilmark{3}
D.\ Fadda,$\!$\altaffilmark{3}
S.\ Laine,$\!$\altaffilmark{3}
F.R.\ Marleau,$\!$\altaffilmark{3}
R.G.\ Carlberg,$\!$\altaffilmark{2}
G.\ Helou,$\!$\altaffilmark{3}
L.\ Yan\,$\!$\altaffilmark{3}}

\email{bridge@astro.utoronto.ca}
\altaffiltext{1}{Some of the data presented herein were obtained at the W.M. Keck Observatory, which is operated as a scientific partnership among the California Institute of Technology, the University of California and the National Aeronautics and Space Administration. The Observatory was made possible by the generous financial support of the W.M. Keck Foundation.}
\altaffiltext{2}{University of Toronto, 50 St. George Street, Toronto, ON, Canada, M5S 3H4}
\altaffiltext{3}{Spitzer Science Center, California Institute of Technology, Pasadena, CA 91125}
\altaffiltext{4}{University of Nottingham, University Park, NG9 2RD, United Kingdom}
\altaffiltext{5}{Department of Physics and Astronomy, Pomona College, CA}

\begin{abstract}
By combining the 0.12 square degree F814W \textit{Hubble Space Telescope} (\textsl{HST}) and \textsl{Spitzer}\ MIPS 24\micron\  imaging in the First Look Survey (FLS), we investigate the properties of interacting and merging Mid-Infrared bright and faint sources at $0.2\le$z$\le1.3$.  We find a marginally significant increase in the pair fraction for MIPS 24\micron\ detected, optically selected close pairs, pair fraction$=0.25\pm0.10$ at z$\sim$1, in contrast to $0.11\pm0.08$ at $z\sim$0.4, while galaxies below our 24\micron\ MIPS detection limit show a pair fraction consistent with zero at all redshifts.  Additionally, 24\micron\ detected galaxies with fluxes $\ge0.1\,$mJy are on average five times more likely to be in a close galaxy pair between $0.2\le$z$\le1.3$ than galaxies below this flux limit.  Using the 24\micron\ flux to derive the total Far-IR luminosity we find that paired galaxies (early stage mergers) are responsible for $27\%\pm9\%$ of the IR luminosity density resulting from star formation at $z\sim1$ while morphologically classified (late stage) mergers make up $34\%\pm11\%$.  This implies that $61\%\pm14\%$ of the infrared luminosity density and in turn $\sim40\%$ of the star formation rate density at $z\sim1$ can be attributed to galaxies at some stage of a major merger or interaction.   We argue that, close pairs/mergers in a LIRG/ULIRG phase become increasingly important contributers to the IR luminosity and star formation rate density of the Universe at $z>$0.7.

\end{abstract}


\section{Introduction}

Hierarchical models and observations suggest that galaxy mergers and interactions play a key role in galaxy assembly and star formation, but to what extent is still unclear.  Studies of gas-rich mergers in the local universe \citep[e.g., Antennae; see][]{sch82} and N-body simulations \citep{mih96,bar04} have revealed fundamental signatures of the galaxy merger process, including tidal tails, multiple nuclei, and violent bursts of star formation.  While interaction-induced star formation is thought to be primarily responsible for ultra luminous infrared galaxies (ULIRGs, which have $L_{IR}\ge10^{12}L_{\sun}$) both locally and at high redshift \citep{san88,das06}, luminous infrared galaxies (LIRGs, $L_{IR}\sim10^{11}-10^{12} L_{\sun}$) appear to have multiple driving mechanisms, merger-induced star formation being only one.  

Luminous infrared (IR) galaxies are thought to be the dominant producers of the cosmic infrared background (CIRB),  and major contributors to the evolution of the cosmic star formation rate (CSFR) of galaxies, especially at $z\ge0.7$ \citep{elb02,lef05}.   The rapid decline from $z\sim1$ of the CSFR density has been linked to a decline in the merger rate. However, recent close pair studies have suggested that the merger rate has remained fairly constant from $z\sim1$ \citep{bun04,lin04}, and at $z\ge0.7$ the IR population is dominated by morphologically normal galaxies \citep{bel05,mel05,lot06}.  The combination of these two results suggest that the bulk of star formation at $z\sim1$ is not driven by major mergers.  

However it must be noted that different merger selection criteria probe different stages of the merger process.  Quantitative measurements of galaxy asymmetry (\citet{abr96a,abr96b,con03}) are more likely to probe later stages, while early stage mergers can be identified by carefully searching for close companions.  There should be some overlap between these techniques if galaxy pairs are close enough to have induced strong tidal interactions, but galaxies in pairs could also have normal morphologies, hence if early stage mergers are not considered, the impact interactions/merging have will be underestimated.

Traditionally, close pair studies have been carried out in the optical/near-IR \citep{pat97,pat00,pat02,car00,lef00,lin04,bun04}.  However recent investigations have begun to explore the Mid-IR properties (star formation) of galaxy pairs, finding a Mid-IR enhancement in pairs separated by less then ten's of kpc's  \citep{lin06}.  The amount of IR luminosity stemming from individual processes (star formation or fueling an AGN) in interacting pairs and mergers still remains open.  To investigate this question we have conducted a study of the frequency of  MIPS 24\micron\ detected,  and undetected close optical galaxy pairs and morphologically defined mergers in the \textsl{Spitzer} First Look Survey (FLS)\footnote{For details of the FLS observation plan and the data release, see http://ssc.spitzer.caltech.edu/fls.}.  We find that the fraction of 24\micron\ detected, optically selected close pairs and mergers increases with redshift, and are important contributors to the IR luminosity and star formation rate density at $z\sim1$. 

In the discussion that follows, any calculation requiring cosmology assumes $\Omega_{\rm M}$=0.3, $\Omega_\Lambda$=0.70, and H$_0$=70\,km\,s$^{-1}$\,Mpc$^{-1}$.

\section{Photometric and Spectroscopic Observations}

The \spitzer\ extragalatic component of the FLS is a 3.7 $deg^{2}$
region centered around R.A.=$17^{h}18^{m}00^{s}$,
decl.=$59^{o}30^{'}00^{''}$.  Observations of this field were taken
using all four Infrared Array Camera (IRAC) channels (Fazio et
al. 2004) and three Multiband Imaging Photometer (MIPS) bands (Rieke
et al. 2004).  Additional ground base images in u*,g' from CFHT's
MegaCam \citep{shi06}, g', i' data from Palomar 200" LFC and NOAO 4-m R
and K' band (Fadda et al. 2004; Glassman et al. 2006 in prep) have also been obtained.  This work
focuses on the 0.12 $deg^{2}$ ACS-HST F814W imaging of the
verification strip, which has 3$\sigma$ depths in MIPS 24\micron\ of
0.1mJy.  Object detection and photometry were performed using
Sextractor (Bertin \& Arnouts 1996).  Particular care was taken to
ensure accurate de-blending of galaxies in close proximity to one
another, while avoiding detections of substructure within a single
galaxy, consistent with other reductions of HST imaging with close
galaxy pairs in mind (Patton et al. 2005).  There were $\sim$59,000
sources extracted within the $F814W_{AB}$ band (hereafter extracted
magnitudes referred to as $I_{AB}$).  We compared our number counts to
those from the Hubble Deep Field (HDF) North and South and determined
a limiting magnitude of $I_{AB}\sim$27.4.

Using the full MIPS catalog from the FLS we selected 24\micron\
sources within the area covered by the ACS imaging ($\sim$0.12
$deg^{2}$).  In order to correlate the MIPS objects with those
identified in the optical we first cross-identified sources from the
MIPS 24\micron\ sample to the IRAC catalog using a tolerance radius of
2.0${''}$.  This choice was primarily motivated by the FWHM of the
MIPS 24\micron\ (PSF$\sim$$6{''}$) and confirmed by visual inspection.
We then cross-correlated the IRAC/MIPS catalog to the ACS sample
which we band merged with u*, g' and $R$ requiring a positional
agreement of $\le$$1{''}$.  When multiple counterparts were identified,
we selected the closest object.  Ultimately we found 1155 ACS sources
also detected by IRAC and MIPS at 24\micron\,.

\begin{figure}[h]
\epsscale{1.2}
\plotone{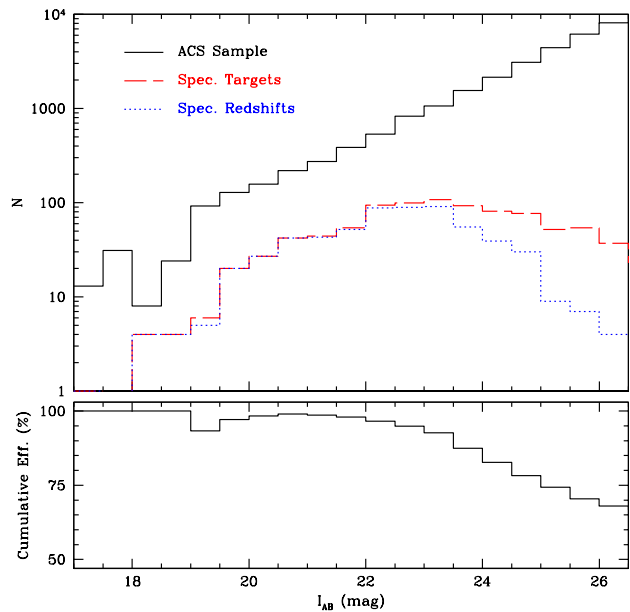}
\caption{({\it Upper}) The $I_{AB}~mag$ distribution for the FLS sample (solid), the sample targeted for spectroscopy (long-dash), and those where spectroscopic redshifts were acquired (dotted).  ({\it Lower}) The cumulative redshift identification efficiency.  The overall spectroscopic completeness is $\sim70\%$ for the full sample and $\sim$80\% for sources with $I_{AB}<25.0$.} 
\label{fig:magdist}
\end{figure}

The redshifts used in this study were determined exclusively from optical
spectroscopy.  They were obtained by cross-correlating the ACS sample,
limited to $I_{AB}\le$26.5 ($N$$\sim$29,000) with various FLS
spectroscopic datasets.  The vast majority of the included redshifts ($\ge$97\%) were obtained with the Deep Imaging Multi-Object Spectrograph (DEIMOS) on the W.M. Keck II 10-m telescope; however, the final sample also included a few redshifts based on Sloan Digitized Sky Survey (SDSS) and WIYN Hydra/MOS (Marleau et al. 2006 in prep) spectra.  Galaxies in the FLS Verification region were targeted for spectroscopic 
follow-up during two DEIMOS campaigns that bracketed Spitzer's launch.  The selection criteria
for these campaigns are summarized below.  For the 2003 pre-launch 
campaign, targets were selected based on NIR ($K_s$) and optical
($g,R,i$) colors.  The primary sample included sources with $K_s$$<$20.2,
$R>$19.0 and a $g,R,i$ color selection that restricted the numbers
of low redshift ($z$$\le$0.6) sources.  For the 2004 post-launch campaign, 
a purely 24\micron\ selected sample ($f_{24}$$>$120\muJy) was targeted for
follow-up.  The combined $I_{AB}$ distribution of targeted and
detected sources is shown in Figure \ref{fig:magdist} ({\it top}) along with the cumulative
redshift identification efficiency ({\it bottom}).  The overall spectroscopic completeness (defined 
here as the fraction of targeted sources with high quality redshifts) is $\sim$70\% for the full sample and $\sim$80\% for sources with $I_{AB}<$25.0.  For a more detailed description of the observing strategy, primary selection criteria and the overall flux and redshift distributions see \citet{cho06}.

Since we were exploring the Mid-IR properties of galaxies no optical limit was imposed, instead an IR luminosity cut ( $L_{IR}\ge 5.0\times10^{10}$ or $1.0\times10^{11}~L_{\sun}$) was used, so that a fair comparison could be made at different redshifts.  The absolute B magnitude ($\rm M_{B}$) distribution of the MIPS spectroscopic sample between 0.5$<$$z$$<$1.3 probes -21$\le$$\rm M_{B}$$\le$-19 fairly uniformly and we are not strongly biased at higher redshifts.

\begin{figure}[h]
\epsscale{1.2}
\plotone{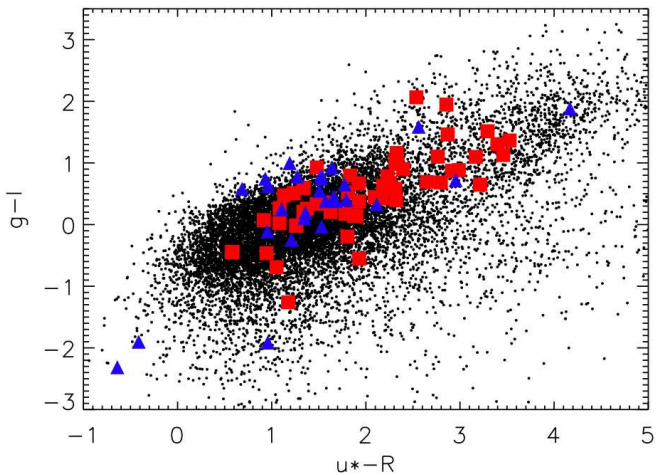}
\caption{Color-color plot where red squares depict the 24\micron\ detected objects with spectroscopic redshifts, while blue triangles show the undetected 24\micron\ spectroscopic sample.  The black dots represent the full FLS-ACS catalog for comparison. The 24\micron\  detected and undetected spectroscopic samples occupy the same color space. Some objects were not detected in all four bands due to the field coverage, depths and seeing differences between filters.}
\label{fig:colors}
\end{figure}

Cross-correlation of the band-merged photometric catalogs with the
redshift samples results in a data set of 476 sources with $I_{AB}<26.5$ and 
$0.2\le$$z$$\le1.3$.  Of those, 245 (51\%) are MIPS 24\micron\,-detected with a
measured $L_{IR}\ge5.0\times10^{10}~L_{\sun}$.  The remaining 231 (49\%)
are non-detected at 24\micron\,.  Despite the fact that the MIPS and non-MIPS galaxies were selected slightly differently, the resultant colors of objects with spectroscopic redshifts have uniform color properties (Figure \ref{fig:colors}).

\vspace*{0.2in}
\section{Statistics of Close Pairs, and Mergers}

To properly constrain the role interactions and mergers play in galaxy evolution, all stages of the process must be considered.    Typically, merger history analyses utilize either pair or structural methods.  Galaxies in pairs are pre-mergers, or systems
undergoing interactions, while morphological or structural methods find galaxies that have already undergone a merger
and are dynamically relaxing.   

When discussing the pair fraction, and inferred merger rates (as defined in Section 4), it must be noted that these measurements are highly dependent on the techniques and selection criteria used to identify ongoing mergers, especially for galaxies at high redshifts. 

The first method, which we will call the ``close pair method", is to count the number of galaxy pairs within some projected separation, $r_{\rm {projected}}$, and magnitude difference ($\Delta m$).   If we assume that these systems will merge within a given time-scale due to dynamical friction, we can determine the merger rate.  Although not all pairs will merge, they can potentially trigger star formation through gravitational interactions.  An alternative is to select merging systems based on morphological indicators either by overall appearance \citep{lef00} or computational measurements such as asymmetry (A), and clumpiness (S) of a system \citep{con00,con02,con03b}, or Gini coefficient (G), and $M_{20}$ parameters \citep{abr03,lot06}.  Due to the comparatively limited spatial resolution of \spitzer\ (compared with optical imaging), seaching for close galaxy pairs or morphological signatures of interaction at Mid-IR wavelengths is currently restricted to the nearby universe. However, we can correlate optically-selected pairs/mergers with global Mid-IR properties and investigate the IR activity in these systems out to high redshifts.

\vspace*{0.4in}
\subsection{Pair Statistics}
We applied the close pairs technique to identify the average number of close companions per galaxy, hereafter $N_{c}$.  This measurement is similar in nature to the pair fraction when there are infrequent triples or higher order N-tuples.  Since this is the case here, $N_{c}$ will be occasionally referred to as the pair fraction.  Companions were selected using a standard operational close pair definition of 5$h^{-1}$kpc$\le r_{\rm{projected}}\le20h^{-1}$kpc, and an optical magnitude difference ($\Delta m$) $\le1.5$ compared to the host galaxy, to select nearly equal mass major mergers.  The term ``host" or ``primary" galaxy are both used to reference the pair member with a measured redshift.  The inner radius is applied to avoid detection of substructure within a galaxy, while the outer $20h^{-1}$kpc limit represents the radius within which satellites are expected to strongly interact with the halo of the host and merge within 0.5-0.9 Gyrs (Patton et al. 1997; Conselice et al 2003).  We find 87 close pairs out of 476 galaxies which fulfill these criteria (see Table 1).  

\begin{figure}[h]
\epsscale{1.2}
\plotone{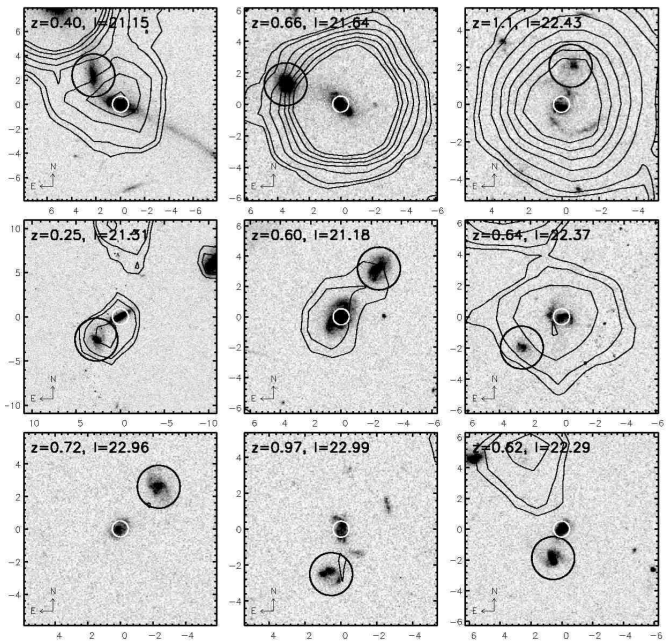}
\caption{A subset of paired galaxies in our sample.  Each image is $60h^{-1}$kpc on a side with axes in arcseconds, centered on the pair member with the spectroscopic redshift also referred to as the primary or host pair member (white circle), while the companion is highlighted in a black circle.  The two upper rows are close pairs which were detected at 24\micron\ , while the lower set are from the undetected 24\micron\ paired sample.  The $3-10\sigma$ 24\micron\ flux contours are overlaid.  The labels are spectroscopic redshift, and $I_{AB}$ magnitude of the primary galaxy.}
\label{fig:stamps}
\end{figure}

To study the fraction of IR-bright galaxies in pairs, we split the pair sample into two sub-sets: those which were detected and those undetected with MIPS at 24\micron\  down to the flux limits of our survey (0.1~mJy).   Figure \ref{fig:stamps} shows a subset of close pairs (both detected and undetected at 24\micron\,) with MIPS contours.  Due to the small separations of close pairs (20$h^{-1}$kpc corresponds to 3.6'' at $z\sim1$) relative to the beam of the MIPS 24\micron\ images (FWHM $\sim$6''), there are a few instances (5) where only a single 24\micron\ detection is found centered between the pair members (see middle left image in Figure \ref{fig:stamps}).  In these cases we assume all 24\micron\ flux is coming from the primary galaxy.  

\subsection{Field Correction}

Since we have redshift information for only the primary galaxy and not the companions we need to consider what fraction of these close pairs are a result of random projection effects.  A field correction was determined using two separate methods to account for these close pairs.  The first assumed the same optical magnitude and redshift distributions independently for both the detected and undetected 24\micron\ samples, while the positions were randomized.  The close pair algorithm was applied to 50 realizations of these mock catalogs and the average $N_{c}$ for each redshift bin was taken to be the pair fraction expected from random.  This assumes the absence of clustering.  

We investigated the environments of 24\micron\ detected and undetected objects on scales of  $r_{\rm{projected}} $$\lsim$20$h^{-1}$kpc, and found them to be comparable, confirming that the increase in pair-fraction of the 24\micron\ detected pairs is not because they preferentially lie in clusters.  On the other hand, there is a weak indication that galaxies detected at 24\micron\ are more likely to lie in small groups. Since such groups may, in some cases, be physical associations, we count such cases as separate pairs.  However, the number of these cases is small,  and does not influence our results in any significant way.  The second method utilizes the $I_{AB}$ magnitude distribution of the full photometric catalog ($\sim$59,000 sources), and determines the average number of companions, within 1.5 mags ($I_{AB}$), normalized to the area covered by  5$h^{-1}$kpc$\le r_{\rm{projected}}\le20h^{-1}$kpc.  The results obtained from the two field correction methods agreed within $\sim2\%$, which is negligible compared to the uncertainly in $N_{c}$.  The average of the two methods was taken to be the final field correction.   Both the pair catalog and randomly generated catalogs were visually inspected for false pairs due to single galaxies being broken up into multiple components in the source extraction phase, or contaminating stars in the photometric catalog, and were removed.  

\subsection{Pair Fractions}

\begin{figure}
\epsscale{1.4}
\plotone{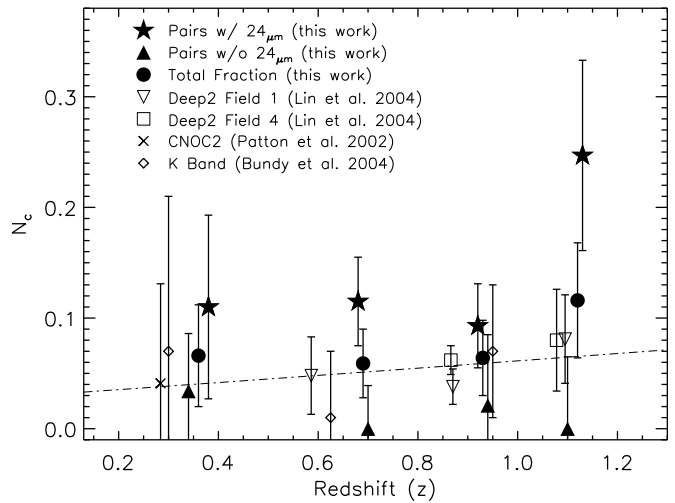}
\caption{ The field corrected pair fraction $N_{c}$ as a function of redshift as measured in the optical and IR.  The stars represent the measurement from our 24\micron\ detected sample,  triangles depict the field corrected pair fraction of the undetected 24\micron\  sub-set, and filled circles show the combined pair fraction of the two samples.  Other optically-determined pair fractions appear as open squares (DEEP2 Field 1) and triangles (Field 2) \citep{lin04}, cross for CNOC2 \citep{pat02}, while the dashed line shows their best power law fit of $(1+z)^{m}$ ($m=1.08\pm0.40$). The near-IR pair fraction determined by \cite{bun04} is shown by diamonds.  Errors for this work are derived using jacknife statistics, while the IR pair fraction errors implemented counting statistics and DEEP2, CNOC2 errors are determined via bootstrap.   A $L_{IR}\ge 5.0\times10^{10}$ limit was imposed on the 24\micron\ detected close pairs.  Note that each work imposes a slightly different luminosity and stellar mass limit. }
\label{fig:pfrac}
\end{figure}

The field-corrected optical pair fractions for the 24\micron\  detected and undetected sub-samples are presented in Figure \ref{fig:pfrac} and Table 1.  Errors are computed using the jackknife technique \citep{efr86}, e.g. given a sample of $N$ galaxies the variance is given by [($N-1)/N \sum_{i}\delta^{2}_{i}$]$^{1/2}$. The partial standard deviations, $\delta_{i}$, are computed for each object by taking the difference between $N_{c}$, the quantity being measured and the same quantity with the ith galaxy removed, $N_{ci}$, such that $\delta{i} = N_{c} -N_{ci}$. 
 
To allow a more direct comparison to be made between the generally lower-luminosity low-z pairs, and those at higher redshift, we derived pair fractions for MIPS detected galaxies with an $L_{IR}\ge5.0\times10^{10} $ (approximately the IR luminosity of the famous Antenna Galaxies). In this way we ensure that the sub-luminous galaxies do not strongly influence the pair fractions in the lowest redshift bin.   
 The derived $N_{c}$ for 24\micron\ detected close pairs is $\sim11\%\pm8\%$ at $z\sim0.4$ and increases to $25\%\pm10\%$ at z$\sim$1.  In contrast, close pairs with no 24\mum\  detection show no increase with redshift and have pair fractions consistent with zero at all redshifts.  The higher pair fraction of MIPS bright sources is marginally significant due to the small number of sources in the highest redshift bin, more MIR selected samples between $z=$1-1.5 are required to strengthen our findings. 

\begin{deluxetable*}{cccccc}
\tabletypesize{\scriptsize}
\tablecaption{FLS Close Pair Statistics}

\tablehead{
\colhead{\textsl{z}} & 
\colhead{\textsl{{$N_{gal}$}}} & 
\colhead{\textsl{{$N_{c}^{D}$}}} & 
\colhead{\textsl{{$N_{c}^{R}$}}} & 
\colhead{\textsl{$N_{c}$}}&
\colhead{\textsl{$\kappa$}}}

\startdata
& &    24\micron\ Detected& & \\
\hline\hfill
 0.2-0.5 &  32 & 0.188 (6) & 0.078 (2.5) & 0.110 $\pm 0.083$ & 0.83 \\
0.5-0.80 & 82   & 0.171 (14)    &  0.057 (4.7)  &  0.114 $\pm 0.040$ & 0.93 \\
0.80-1.0 & 82 & 0.122 (10) & 0.029 (2.4) & 0.093 $\pm 0.038 $  & 0.90\\
1.0-1.3  &   49   & 0.429 (21) & 0.182 (8.9) & 0.247 $\pm 0.086$ & 0.67\\
\hline
& &    24\micron\ Undetected& &\\
\hline
0.2-0.5 &   44 & 0.136 (6) & 0.102 (4.5) & 0.034$ \pm 0.052 $ & 1.00\\
0.5-0.80 & 76   & 0.132 (10)    &  0.134 (10.2)  & 0$\pm 0.039$ & 1.00  \\
0.80-1.0 & 56 & 0.214 (12) & 0.193 (10.8) & 0.021$\pm 0.064$ & 0.83 \\
1.0-1.3 &  55   & 0.145 (8) & 0.180 (9.9) & 0 $\pm 0.065$ & 0.75 \\
		      
\enddata
\tablecomments{\textsl{$N_{gal}$} is the number of galaxies with a spectroscopic redshift, \textsl{$N_{c}^{D}$} is the number of companions per host fulfilling our pair criteria, while \textsl{$N_{c}^{R}$} is the number of projected companions per host from the field.  The corrected fraction of companions per host is given as, (\textsl{$N_{c}^{D}$}-\textsl{$N_{c}^{R}$})/\textsl{$N_{gal}$}, with errors being determined using a jacknife technique. Numbers appearing in parentheses refers to the number of close pairs in the respective redshift bins.  Undetected at 24\mum\ refers to sources below the limits of our survey ($0.1\,$mJy).  The constant $\kappa$ is the fractional number of mergers per host galaxy.  A $L_{IR}\ge 5.0\times10^{10}$ limit was imposed.}
\end{deluxetable*}
 
We would like to be able to rule out the possibility that $N_{c}$ is biased by the brightest IR sources at $z\ge1$, since merger fractions change as a function of luminosity and mass \citep{con03,Xu04}.  To address this we placed a higher IR luminosity limit ($L_{IR}\ge 7.0\times10^{11}$) on the sample, so that at $z\ge0.7$ the same populations were being probed (optically we are probing -22$\lsim$$\rm M_{B}$$\lsim$-19).  We still find an increase in $N_{c}$ from the lower ($0.8\le z \le 1.0$) to the higher ($z\ge1$) redshift bins of similar magnitude compared to when the lower IR limit ($L_{IR}\ge 5.0\times10^{10}$) was used.  Therefore the increase in $N_{c}$ found at $z\ge1$ is likely not a result of merely probing brighter IR systems but rather due to a physical increase in the merge rate for the 24\micron\ population, however deeper 24\micron\ imaging and spectroscopy are required to confirm this.  When we consider the averaged pair fraction over $0.2\le z \le 1.3$ for the 24\micron\ detected sample we find that galaxies above a flux limit of $0.1\,$mJy are five times more likely to be in a close galaxy pair, than those below this limit. 

\subsection{Morphological Mergers}

To explore the structural components of galaxies in our sample we used the CAS (Concentration, Asymmetry, Clumpiness) quantitative classification system \citep{con97, con00, con02,con03b}, and visual classifications.  To measure the merger fraction using structural classifications we visually inspected the full 24\micron\  detected spectroscopic catalog with the following groupings: early type (E, S0), mid-types (Sa-Sb), late-types (Sc-irr), compact systems, disturbed disks, and mergers.  The methodology for carrying out this classification is described in detail in \citet{con05}.  Basically, each galaxy was viewed on a computer screen and classified into one of our types.  Overall we find that $55\%\pm5\%$ of 24\micron\  detected galaxies are disks, which is consistent with \citet{bel05,lot06}, while $26\%\pm5\%$ are merging systems and $\sim6\%$ were classified as disturbed disks and are possible minor mergers.  A fraction of the disk-dominated objects do show some visual signs of a morphological disturbance, or are in a pair, as we will discuss later in this paper.

\begin{figure}
\epsscale{1.20}
\plotone{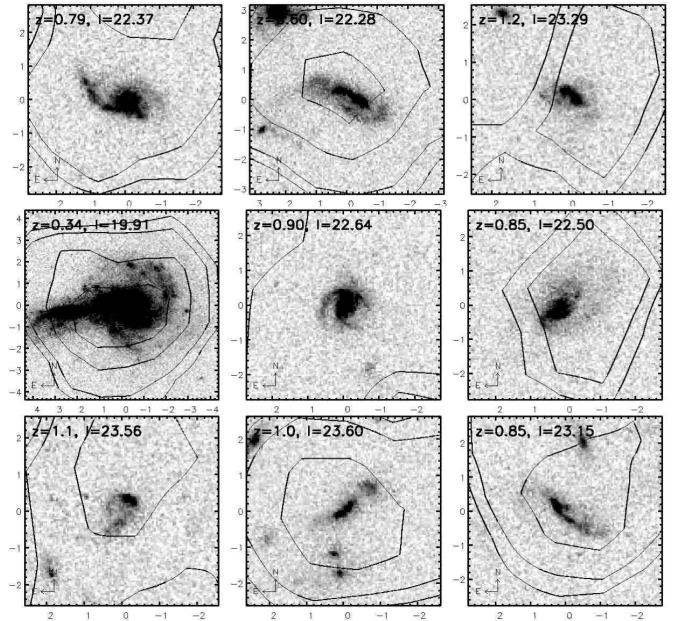}
\caption{A subset of 24\micron\ detected galaxies in our sample classified as a merger.  Each image is $30h^{-1}$kpc on a side with axes in arcseconds.  The $3-10\sigma$ 24\mum\ flux contours are overlaid.  The labels include spectroscopic redshift, and $F814W_{AB}$ magnitude.}
\label{fig:stamps2}
\end{figure}

\begin{figure}
\epsscale{1.3}
\plotone{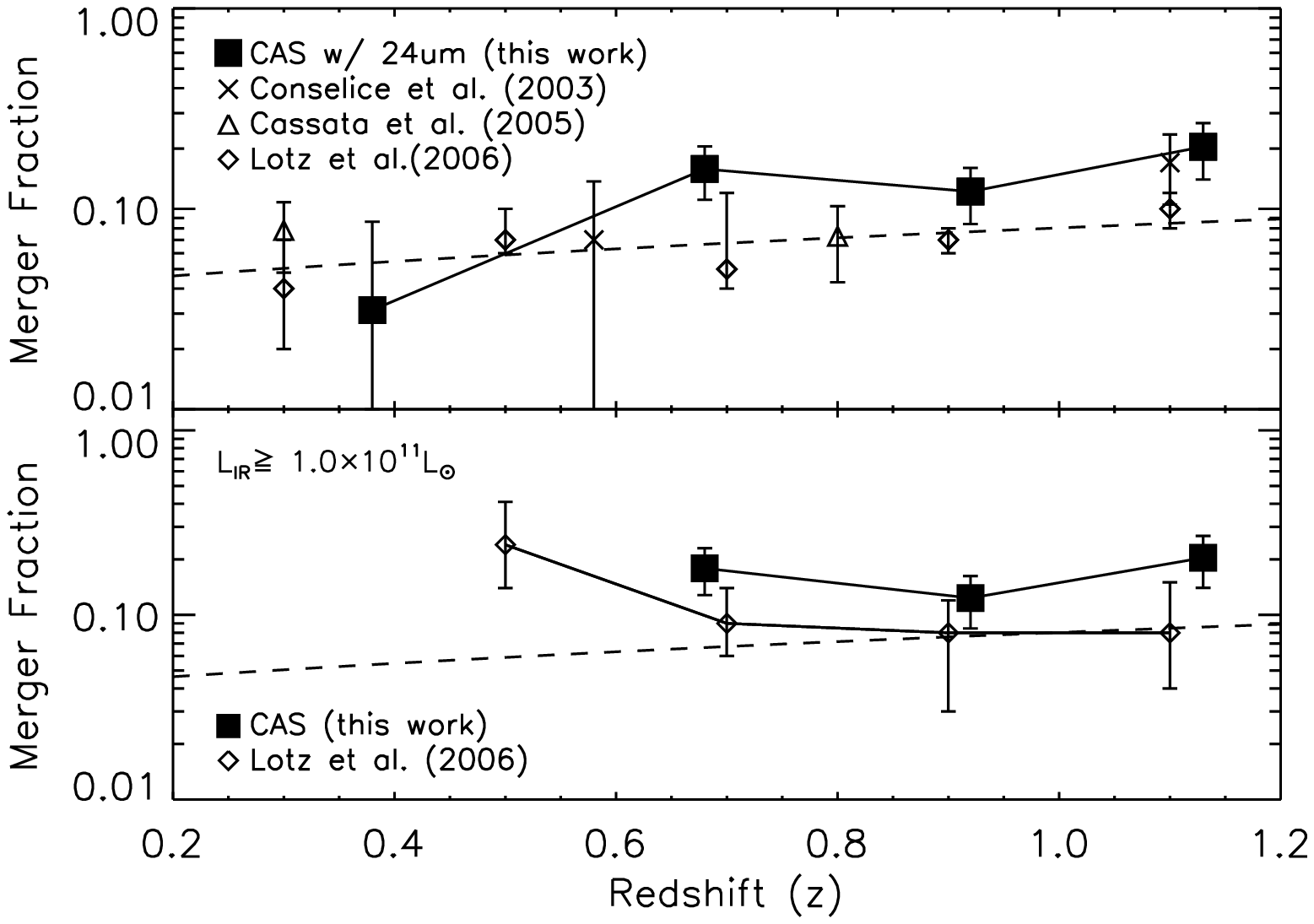}
\caption{Merger fraction as a function of redshift, using quantitative morphological criteria.  The filled squares represent measurements from FLS (this work) of sources with a 24\micron\ detection.  The``x's" mark the results of \citet{con03}, triangles depict \citet{cas05}, and diamonds represent \citet{lot06}.  Top panel shows the merger fraction of other studies with no 24\micron\ criteria imposed, while the measurements from this work are mergers with L$_{IR}\geq 5.0\times10^{10}$L$_{\sun}$.  The dashed lines shows the best fit of $(1+z)^{m}$ using all points (from the other studies) with no MIPS limit imposed ($m=1.08$). Bottom panel shows the merger fraction for LIRG/ULIRG galaxies (L$_{IR}\geq 1.0\times10^{11}$L$_{\sun}$).  Error bars derived using Poisson statistics.}
\label{fig:mfract}
\end{figure}

Galaxies undergoing a major merger event can also generally be identified by their large asymmetries in the rest frame optical \citep{con00, con03}.  We defined a major merger as a galaxy having an asymmetry (A) $\ge\,0.35$ and $I_{AB}\,\le\,26.5$ (see Figure \ref{fig:stamps2} for examples).  This limit has been shown to be a clean way to find galaxy mergers, without significant contamination from non-merging galaxies \citep{con03b}. Figure \ref{fig:mfract} shows how the merger fraction for CAS defined mergers evolves as a function of redshift for both 24\micron\ detected objects (top panel) and LIRG/ULIRG galaxies (bottom panel).   As with the 24\micron\ detected close pair sample there is an elevated merger fraction compared to other works \citep{cas05,lot06} in which no 24\micron\ flux limit was imposed, and a slight indication of evolution with redshift, but it is statistically consistent with $m\sim1.0$ (dashed line), where $m$ is the slope of a power-law of form $(1+z)^{m}$ later discussed in $\S5$.  

\begin{figure}[h]
\epsscale{1.2}
\plotone{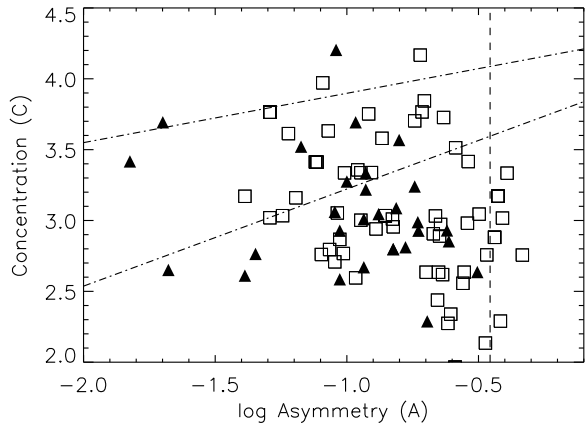}
\caption{Asymmetry - Concentration diagram for the 24\micron\ detected close pairs (squares) and undetected pairs (triangles).  The long dashed (vertical) line separates merging and non-merging systems (A$\ge0.35$ is considered a merger), while the dash-dotted lines separate early (upper) to mid to late (lower) type galaxies defined by \citet{con03b}. Generally close pairs are mid/late type galaxies, and some would also be morphologically classified as a merger.}
\label{fig:cas}
\end{figure}

 We also performed a CAS analysis of our close pairs sample which revealed that 24\micron\ detected pairs are notably more asymmetric than the undetected-MIPS close pairs (Figure \ref{fig:cas}), suggesting that interactions and collisions may play a role in their IR activity.   If the 24\micron\ detected close pairs were generally of a different morphological classification than those pairs undetected at 24\micron\, the discrepancy in the asymmetries could be explained.  To address this issue each close pair was visually inspected and classified by four of the authors to be either disk or bulge-dominated.  We find that 81\% of the 24\micron\ pairs have disk morphologies while $74\%$ of the undetected 24\micron\ hosts were also disk dominated, hence the discrepancy between the asymmetries of the two groups is not caused by classification differences, but rather is a physical effect.
 
\section{Merger Rates}

One of the goals of studying mergers and interactions is to determine how the galaxy merger rate evolves with redshift.  Most studies of galaxy mergers involve determining the merger fraction, yet the merger rate, which is defined as the number of galaxies merging per unit time per unit volume, is a more physical quantity that can be used to determine the full merger history. The rate in which galaxies merge also affects the mass function of galaxies, and is likely linked to the cosmic star formation rate.   Since we are considering a very broad range in the merger process, from early-stage or pre-mergers selected via close galaxy pairs, and later-stage mergers chosen based on morphological criteria, we must be careful when determining their respective merger rates, as the time-scales for these processes are all different.  

There are two variations of the merger rate definition.  The first is the number of mergers that a galaxy will undergo per unit time ($\Re_{mg}$), and the second is the total number of mergers taking place per unit time per unit co-moving volume ($\Re_{mgv}$).  Since we are primarily interested in mergers which are also Mid-IR bright systems we will have to restrict ourselves to measuring $\Re_{mg}$ because the evolution of the 24\micron\ luminosity function with redshift is currently not well constrained, and our redshifts are not complete enough to reconstruct this evolution.

In order to determine $\Re_{mg}$ we need to identify systems which are destined to merge.  We have approached this measurement from three different perspectives, close pairs to select pre-mergers or interactions, visual inspection to select interactions after the first passage, and late stage mergers, as well as CAS criteria which quantitatively selects for later stage mergers.  By combining information about the number of ongoing mergers ($N_{m}$) and the time-scales, ($T_{mg}$) on which they will undergo said merger, one can estimate an overall merger rate $\Re_{mg} = N_{m}/T_{mg}$.  Each method of identifying mergers/interactions is capturing a different snapshot of the merger process, each with different merger timescales.  

The value of $N_{c}$ is directly proportional to the number of mergers per galaxy ($N_{m}$), such that $N_{m} = \kappa N_{c}$ ($\kappa$ is a constant relating to the number of mergers per galaxy).  Hence, the merger rate detemined using close galaxy pairs is given by $\Re_{mg} = \kappa N_{c}/T_{mg}$.  The value of $\kappa$ depends on the nature of the merging systems under consideration.  If one were to identify a pure set of galaxy pairs each consisting of one companion undergoing a merger, then $\kappa=1.0$.  In our case it exclusively accounts for close pairs which are in doubles and perhaps higher order N-tuples.  Our definition of $\kappa$ differs by a factor two from \citet{pat00} which in this instance would have $\kappa=0.5$ since they have redshifts for both pair members and one merger is made up of two companions.  We have redshift information for only one pair member, therefore one merger is made up of a primary and one companion. The merger rate equation for merging galaxies selected by visual classification and CAS parameters is simply $\Re_{mg} = f_{mg}/T_{mg}$, where $f_{gm}$ is the galaxy merger fraction.  

Before the merger fraction can be used to calculate the merger rate we need to understand the time-scale in which a merger occurs.   Each technique of identifying mergers has a different time-scale since each is sensitive to a different interval of the merger process.  There are two main methods that have been used to estimate the time-scale of a merger: dynamical friction arguments, and N-body models.  The details of these methods are beyond the scope of this paper but see \citet{pat00} and \cite{con06} for a review.  We take the average merger time-scale for a set of close companions of roughly equal mass to merge as $\sim$0.5 Gyr$\pm0.25$, derived from dynamical arguments \citep{pat00,con06}.  \citet{con06} showed through N-body simulations that visual classification selects on-going mergers over a longer time-scale (1.0 Gyr$\pm0.25$) since the human eye detects both early and later stage mergers, while the asymmetry of a galaxy is sensitive to 0.41 Gyr$\pm0.17$ \citep{con06} of the merger sequence.  

\vspace*{0.4in}
\section{The Evolution of the Galaxy Merger Rate $0.2\le z \le 1.3$}

Within the past two decades numerous studies have been performed to estimate the evolution of the galaxy merger fraction, using both the close pair technique \citep{zep89,bur94,car94,yee95,pat97,pat00,lef00,lin04,bun04} and morphological parameters \citep{lef00,con03,lav04,lot06}.   Evolution in the galaxy merger fraction is often parameterized by a power-law of form $(1+z)^{m}$, and has yielded a wide range of results, spanning 0$\la$$m$$\la$5.  The large spread in values is in part due to the different selection criteria used to identify merging systems and biases from optical contamination or redshift completeness.  \citet{pat97} considered these biases and demonstrated that most results to that date were consistent with their estimate of $m=2.9\pm0.9$.  Recently, optical and near-IR close pair studies \citep{lin04,bun04} have derived merger fractions with little redshift evolution ($m$$\sim$1), as have some morphological studies using ($G$), and $M_{20}$ \citep{lot06}.    

\begin{figure}[h]
\epsscale{1.2}
\plotone{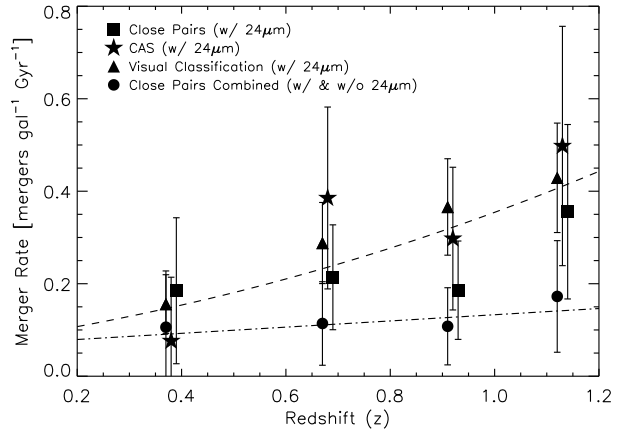}
\caption{The number of mergers per galaxy ($L_{IR} \ge 5\times 10^{10}$) as a function of redshift. Three merger/interaction selection techniques are applied, close pairs (squares), CAS criteria (stars), and visual classification (triangles), while merger rates for the combined (w/ and w/o 24\micron\ detections) using the close pairs method are shown with circles.  The long dashed curve is the best fit of the form $(1+z)^{m}$, using the FLS data for the three techniques; the dot-dashed curve represents the best fit for the combined total close pairs (MIPS and non-MIPS pairs).  }
\label{fig:mr}
\end{figure}

When we consider all the close pairs identified in our sample,  both those detected at 24\micron\ and not, we find a merger fraction and rate consistent with recent studies showing little redshift evolution.  However, when we separate the pair sample into systems with a 24\micron\ detection above 0.1 mJy, and those below it, we do see a stronger evolution of $N_{c}$  with redshift, (recall that $N_{c} \propto \Re_{mg}$) and therefore also in the merger fraction and rate (Figure \ref{fig:mr}).  Similarly, visually classified mergers and those identified via asymmetry levels ($A\ge0.35$) using the CAS parameters, also show redshift evolution in the merger fraction and rate.  The merger fraction computed using the different methods are in good agreement when normalized by their respective time-scales, reinforcing the idea that we are probing different phases of the merger process.  Considering all three merger selection techniques we find the best fit of the merger rate parameterized by $\Re(0)(1+z)^{m}$ to be $0.077 \pm 0.045$, 2.12 $\pm0.93$, with a reduced $\chi^{2}$=0.39.   This result suggests that when one considers a sample of close galaxy pairs solely on their optical fluxes, brighter than $\rm M_{B}=$-19, little evolution of the merger rate with redshift is found.  However close pairs emitting 24\micron\ flux exhibit an increase in the merger rate with redshift.  The infrared luminosity limit ($L_{IR}\ge10^{11}L_{\sun}$) imposed on the close pairs and mergers allows us to primarily probe systems in a LIRG/ULIRG phase at $z\ge0.4$ (see next section for details).   The increase of the merger fraction and rate of this population of galaxies coupled with the fact that LIRG/ULIRG galaxies dominate the SFR density at $z\ge0.7$ \citep{lef05} suggests that merging does in fact play an increasingly important role in star formation out to $z\sim1$.

\section{Total Infrared Luminosities of Mergers}

\begin{figure}[h]
\epsscale{1.25}
\plotone{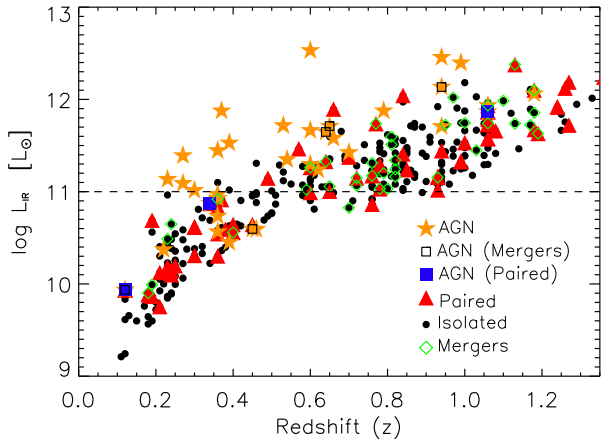}
\caption{ Infrared luminosity $L_{IR [8-1000\mu m]}$ vs. redshift for the MIPS 24\mum\ sample with spectroscopic redshifts, broken down into galaxies in a close pair (red triangle), CAS mergers (open green diamonds), AGN candidates (yellow star), AGN candidates in a close pair (blue square), AGN mergers (open black square), and field galaxies (black circles).  LIRGs lie above the horizontal dashed line at $L_{IR}\ge10^{11}L_{\sun}$}
\label{fig:lirall}
\end{figure}

One way to quantify the role merging galaxies play in triggering star formation is to investigate their contribution to IR luminosity densities.  Infrared luminosities (8-1000 \micron\ ) were calculated utilizing the 24\micron\ fluxes and two different template methods: \citet{cha01,dal01} in a similar manner as \citet{lef05} for the full MIPS 24\micron\  spectroscopic sample (Figure \ref{fig:lirall}). MIPS pairs and mergers share a similar luminosity distribution to 24\micron\ bright field galaxies, although red-AGN seem generally more luminous which is in part due to template mismatches \citep{cha01}.  

\begin{figure}[h]
\epsscale{1.20}
\plotone{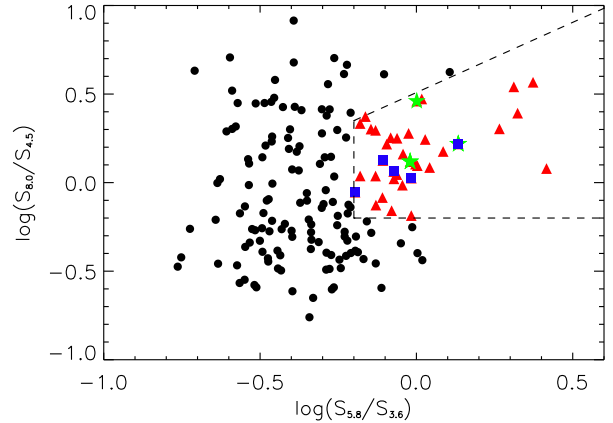}
\caption{ IRAC color-color plot using the region in the FLS covered by the ACS imaging.  Circles represent objects with spectroscopic redshifts and detections in all four IRAC channels. The red triangles indicate objects which have met the color criteria (shown by the dashed line) of an AGN candidate \citep{lac04}.  The green stars and blue squares depict objects in a close galaxy pair or merger whose host was also flagged as an AGN candidate. }
\label{fig:agn}
\end{figure}

The $L_{IR}$ of a galaxy is a combined measure of the reprocessed UV photons intercepted by dust from massive young stars and AGN.  Therefore to investigate the contribution an interacting or merging galaxy makes towards the total $L_{IR}$ density from star formation alone we must first remove AGN from our sample.  Due to the nonuniform rest-frame spectral coverage of our sample we rely on the four-band IRAC color selection used by \citep{lac04} to identify and remove AGN candidates (Figure \ref{fig:agn}).  Over the modest redshift range of our sample, this method is still effective at separating IR-warm AGN from  starburst systems.  We find an AGN contamination rate of $\sim12\%$ for the full 24\micron\ sample, while $\sim14\%$ of the hosts in a pair or merger were characterized as AGN. 

With AGN candidate objects removed we can infer the contribution to the $L_{IR}$ density from star formation coming from 24\micron\ galaxies in a interaction/merger as a function of redshift.   We derive the number of statistically ``real" galaxy pairs from our pair fraction result at each redshift interval and determine the total $L_{IR}$ density from close pairs which is in turn divided by the $L_{IR}$ density from the whole sample.  We find that paired galaxies ($L_{IR}\ge10^{11}L_{\sun}$) are responsible for $27\%_{-8\%}^{+9\%}$ of the IR background stemming from star formation at $z\sim1$.  Since we only know the redshift of the host galaxy we select ``real" close pairs in a statistical sense, and derive error bars for the close pairs contribution by the spread of 50 realizations of the $L_{IR}$ density from different combinations of 24\micron\ galaxy pairs.   We also applied this analysis to CAS and visually classified mergers, which make up an additional $\sim$12\%, and $\sim$22\% of the IR luminosity density respectively.  Naturally there is a some overlap in mergers identified through close pair criteria and morphological parameters, since interacting pairs can exhibit tidal tails and asymmetric structures, causing them to also be identified morphologically as mergers.   We found that $37\%$ of CAS defined mergers were also in a close pair, and 31\% of visually identified mergers were also classified by CAS as merging.  In cases where a merging system was identified using multiple techniques it's contribution was only counted once.  For example, if  a merger identified morphologically (either through CAS or visual inspection) is also in a close pair it is removed from the morphological merger catalog, or if a CAS merger is also identified visually the merger is removed from the visual merger catalog.  This insures that no close pair or merger is counted more than once when deriving the contribution from interactions and mergers to the IR luminosity density.

\begin{figure}[h]
\epsscale{1.2}
\plotone{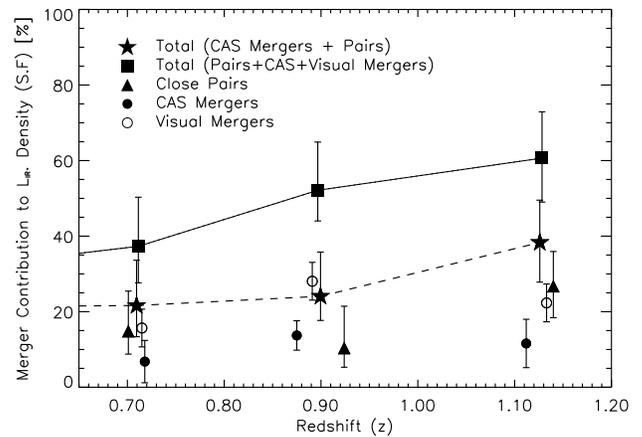}
\caption{The fraction of $L_{IR}$ density (as a result of star formation) as a function of redshift coming from LIRG/ULIRG galaxies in a close pair (pre-merger phase) shown by triangles, or more advanced stage mergers defined morphologically (circles).  The star symbol indicates the total combined contribution from close pairs and CAS mergers, while squares depict the total from close pairs, CAS, and visually classified mergers. Note an infrared limit of $L_{IR}\ge10^{11}L_{\sun}$ was imposed.}
\label{fig:lumden}
\end{figure}

The combination of these three merger selection techniques identifies a large range in the merger process, from pre-merger to late stage mergers, implying that $\sim$60\% of the infrared luminosity density at $z$$\sim$1 can be attributed to galaxies involved in some stage of a major merger (Figure \ref{fig:lumden}).  The remaining $\sim$40\% of the IR background from LIRG/ULIRGs is likely to predominately come from active, isolated gas-rich star-forming spirals, with some contribution from minor mergers.  

If we exclude visually classified mergers the close pair/merger contribution to the IR density is $\sim$38\%, in good agreement with \citet{lin06} who estimate a moderate contribution from interacting and merging systems of $\lsim$36\%.  It must be noted however that neither \citet{lin06} or this work have considered the contribution from minor mergers and are therefore lower limits. 

\vspace*{0.2in}
\subsection{Star Formation in Mergers \& Interactions}

An important and highly debated question is: how important are galaxy mergers in understanding the dramatic decline of the cosmic SFR density from $z\sim1$ to the present day?  It has been well established that mergers and interactions can induce violent bursts of star formation \citep{sch82,bar00,mih96,cox06}.  So to investigate this contribution we derived the SFR for our 24\micron\ detected close pairs and mergers, using their $L_{IR}$.  The infrared luminosity of a galaxy is a star formation rate tracer which is unaffected by the extinction of dust.  The dominant heat sources of most dusty, high-opacity systems such as LIRGs and starbursts is stellar radiation from young stars.  In these types of systems the $L_{IR}$ can be converted into a SFR using the calibration of \citet{ken98}, $SFR_{IR}=4.5 \times 10^{-44}L_{IR}(\rm{ergs\, s}^{-1})$, where $L_{IR}$ is the integrated luminosity from 8-1000\micron\ as determined in section 6.0.  

We estimated the contribution mergers and interactions above $L_{IR}\ge10^{11}L_{\sun}$ make to the SFR density at $z\sim$1 in two ways.  The first is simply to consider their contribution to the $L_{IR}$ density which is a star formation tracer.  Section 6.0 determined that mergers and interactions at z$\sim$1 (above $L_{IR}\ge10^{11}L_{\sun}$) are responsible for 40-60\% of the IR luminosity density.  Using the results of \citet{lef05}  which showed that $z\ge$0.7 LIRGs produce $\sim70\%$ of the star formation rate density, we can infer that mergers and interactions in LIRG/ULIRG phase would be responsible for $\sim30-40\%$ ($0.6\times70\%$) of the SFR density at $z\sim$1, since IR activity traces dusty star formation.   

The second more detailed approach utilizes the SFR density directly arising from our sample of mergers and interactions.   At 1.0$\le$$z$$\le$1.3 we find that 59\% (12 close pairs, and 17 later stage mergers) of galaxies detected at 24\micron\ are involved in some stage of an interaction or merger.   In this same redshift range our sample is insensitive to galaxies with IR luminosities $\le$$10^{11.5}$.  To correct for this we derived a scaling factor ($\sim$7) simply by comparing the number of observed objects of a given $L_{IR}$ in a specific redshift range to the number expected from models \citep{lag04}.  However to go any further we must assume that our spectroscopic sample is representative of this population at $\sim$1, and by all accounts this appears to be true.   Using the derived pair fraction we can then infer the total number of major mergers and interactions occurring (fulfilling our criteria) in a given volume and $L_{IR}$ limit.  The lower limit of the SFR density at $z\sim$1 from merging and interacting galaxies is found to be 0.066 $\rm{M}_{\sun} \rm{yr}^{-1} \rm{Mpc}^{-1}$.  Using the extinction corrected ``Lilly-Madau" plot (0.1585 $\rm{M}_{\sun} \rm{yr}^{-1} \rm{Mpc}^{-1}$ at $z=$1) \citep{tho01} we find that mergers and interactions are responsible for at least $42\%$ of the SFR density at $z\sim$1 (assuming mergers contribute 60\% of the IR density).  Both approaches are in good agreement, and are only a lower limit, since objects flagged as AGN were not considered even though some of their $L_{IR}$ is a result of star formation, and minor mergers which have been shown to also induce bursts of star formation were not included.

These results have interesting implications for the physical mechanisms that drive the decline in the cosmic SFR (CSFR) density from $z\sim1$ to present day.   They suggest that when all stages of the merger process are considered (pre-merger to later stage merger) major interactions and mergers contribute close to half of the $z\sim1$ SFR density, and the decline in the number of 24\micron\ detected mergers/inteactions is a significant,  but perhaps not the primary driver for the decline in the cosmic SFR.  This conclusion differs in interpretation from \citet{bel05,mel05,wol05,lin06,lot06}, which generally suggest that the evolution of the merger rate is not a significant underlying cause of the decline in the cosmic SFR, but rather a strong decrease in the SFR of morphologically undisturbed spiral galaxies is the dominant mechanism.   Their results do not preclude the possibility that their ``star forming (undisturbed) disks" could be in widely separated pairs, and when we only consider quantitatively defined morphological mergers our results are consistent with theirs stressing the importance of considering the merger process in its entirety.   It must also be mentioned that we are probing to higher redshifts than \citet{bel05}, which found that major galaxy mergers account for $\le$30\% of the IR luminosity density at $z$$\sim$0.7, consistent with our findings of 35\% at that redshift.  Our results also agree that at $z\sim$0.7 isolated undisturbed spiral galaxies are a primary contributor, however, the influence shifts to interactions and mergers at $z$$>$0.7. 

Our findings point to an increased importance of MIPS bright interactions and mergers to the IR luminosity density and SFR density at $z$$\ge$0.7.  This conclusion is not hampered by the small statistics of the $z>$1 bin.  Figure \ref{fig:lumden} shows the IR luminosity density contribution from interactions/mergers at $z$$\sim$0.7 to be $\sim$37\% and 52\% at $z$$\sim$0.9, reinforcing this increasing trend.  

\vspace*{0.1in}
\section{Discussion}

Using a spectroscopic sample of field galaxies from the ACS component of the FLS and dividing it into two subsets, those with a 24\micron\ detection (above $0.1\,$mJy) and those without (or below) we identified optically merging/interacting systems via close pair statistics and morphological methods.   We find that roughly 25\% of galaxies emitting at 24\micron\ have a close companion at $z\sim$1 while at $z\sim$0.5 only $\sim11\%$ are in pairs.  In contrast, those undetected at MIPS 24\micron\ showed a pair fraction consistent with zero at all redshifts (0.2$\le$$z$$\le$1.3).  On average MIPS 24\micron\ galaxies are five times more likely than non-MIPS sources to have a close companion over 0.2$\le$$z$$\le$1.3.   When the samples are combined (regardless of 24\micron\ flux) we find pair fractions consistent with previous studies \citep{lin04,bun04} showing little evolution with redshift.  

 An important and open question is the cause of star formation in LIRG galaxies at high-z.  Some morphological studies have suggested that since at least half of the LIRG galaxies exhibit disk dominated morphologies \citep{bel05,lot06} at $z\ge$0.7 and low non-evolving merger fractions \citep{lot06} that the driver of IR activity in high-z LIRGs is from on-going star-formation from isolated gas-rich spirals and not merger or interaction induced.  One bias of $\textsl{HST}$ morphological studies involving the identification of merging/intereacting systems is the limitation of detecting low surface brightness features such as tidal tails caused by close interactions, which can lead to an underestimate of the importance of mergers in the evolution of galaxies at $z<$1.  Ultimately both close pair and morphological  techniques must be applied and considered, to obtain a complete major merger timeline.  Our analysis is the first to probe merger rate evolution combining close pairs and later stage mergers while considering the IR activity of these systems.  
 
We find that close pair statistics, visually classified mergers, and those identified via quantitative CAS parameters all showed similar evolution in their merger rates.  Fitting the merger rate evolution function $\Re(z) \propto (1+z)^{m}$ for 24\micron\ detected mergers above $0.1\,$mJy, we find $m=2.12 \pm0.93$.  This result agrees with previous claims of an increase ($m\ge2$) of the merger rate out to $z\sim1$ \citep{pat97,pat00,lef00,con03,cas05}.  However this evolution is not seen when IR faint ($<0.1\,$mJy) mergers are included, suggesting that it is the LIRG-merger population that is evolving with redshift.  
 
The Mid-IR emission of LIRGs is indicative of dust enshrouded star formation (and some AGN activity), and at $z\ge0.7$ they dominate the IR luminosity density and in turn the volume-averaged star formation rate density at $z\sim1$.   We estimate that close galaxy pairs are responsible for $\sim27\%$ of the IR luminosity density resulting from star formation at $z\sim1$, while later stage mergers contribute $\sim35\%$.  This implies that 40-60\% of the infrared luminosity density at $z\sim1$ can be attributed to galaxies involved in some stage of a major merger, indicating that merger-driven star formation is responsible for 30-40\% of the star formation density at $z\sim1$.  This value is a lower limit since minor mergers and interactions/mergers with an AGN were not considered.
 
Ultimately, our findings suggest that interactions and mergers of LIRG phase galaxies play an increasingly important role in both the IR luminosity and SFR density from $z\ge0.7$ out to $z\sim1.3$, and are vital to our understanding of the evolution and mass assembly of luminous IR galaxies.

\acknowledgments
We would like to thank H. Shim, M. Im, R. Chary, and C. Borys for their contributions to this work, V. Charmandaris for useful suggestions, and the anonymous referee for valuable comments that improved the clarity of the paper.  This work is based on observations made with the \textsl{Spitzer Observatory}, which is operated by the Jet Propulsion Laboratory, California Institute of Technology, under NASA contract 107.  Support for this work was provided in part by the \textsl{Spitzer Graduate Student Fellowship} program and an Ontario Graduate Scholarship in Science and Technology.   The authors wish to recognize and acknowledge the very significant cultural role and reverence that the summit of Mauna Kea has always had within the indigenous Hawaiian community.  We are most fortunate to have the opportunity to conduct observations from this mountain.


\begin{thebibliography}{}
\bibitem[Abraham et al.(1996a)]{abr96a} Abraham, R.~G., van den Bergh, S., Glazebrook, K., Ellis, R.~S., Santiago, B.~X., Surma, P., \& Griffiths, R.~E.\ 1996, \apjs, 107, 1
\bibitem[Abraham et al.(1996b)]{abr96b} Abraham, R.~G., Tanvir, N.~R., Santiago, B.~X., Ellis, R.~S., Glazebrook, K., \& van den Bergh, S.\ 1996, \mnras, 279, L47 
\bibitem[Abraham et al.(2003)]{abr03} Abraham, R.~G., van den Bergh, S., \& Nair, P.\ 2003, \apj, 588, 218
\bibitem[Barnes(2004)]{bar04} Barnes, J.~E.\ 2004, \mnras, 350, 798 
\bibitem[Barton et al.(2000)]{bar00} Barton, E.~J., Geller, M.~J., \& Kenyon, S.~J.\ 2000, \apj, 530, 660
\bibitem[Bell et al.(2005)]{bel05} Bell, E.~F., et al.\ 2005, \apj, 625, 23 
\bibitem[Bertin \& Arnouts(1996)]{ber96} Bertin, E., \& Arnouts, S.\ 1996, \aaps, 117, 393 
\bibitem[Bundy et al.(2004)]{bun04} Bundy, K., Fukugita, M., Ellis, R.~S., Kodama, T., \& Conselice, C.~J.\ 2004, \apjl, 601, L123
\bibitem[Burkey et al.(1994)]{bur94} Burkey, J.~M., Keel, W.~C., Windhorst, R.~A., \& Franklin, B.~E.\ 1994, \apjl, 429, L13
\bibitem[Carlberg et al.(1994)]{car94} Carlberg, R.~G., Pritchet, C.~J., \& Infante, L.\ 1994, \apj, 435, 540 
\bibitem[Carlberg et al.(2000)]{car00} Carlberg, R.~G., et al.\ 2000, \apjl, 532, L1 
\bibitem[Cassata et al.(2005)]{cas05} Cassata, P., et al.\ 2005, \mnras, 357, 903 
\bibitem[Chary \& Elbaz(2001)]{cha01} Chary, R., \& Elbaz, D.\ 2001, \apj, 556, 562
\bibitem[Choi et al.(2006)]{cho06} Choi, P.~I., et al.\ 2006, \apj, 637, 227 
\bibitem[Conselice(1997)]{con97} Conselice, C.~J.\ 1997, \pasp, 109, 1251
\bibitem[Conselice et al.(2000)]{con00} Conselice, C.~J., Bershady, M.~A., \& Jangren, A.\ 2000, \apj, 529, 886 
\bibitem[Conselice et al.(2002)]{con02} Conselice, C.~J., Gallagher, J.~S., III, \& Wyse, R.~F.~G.\ 2002, \aj, 123, 2246 
\bibitem[Conselice(2003)]{con03b} Conselice, C.~J.\ 2003, \apjs, 147, 1 
\bibitem[Conselice et al.(2003)]{con03a} Conselice, C.~J., Bershady, M.~A., Dickinson, M., \& Papovich, C.\ 2003, \aj, 126, 1183 
\bibitem[Conselice et al.(2003)]{con03} Conselice, C.~J., Chapman, S.~C., \& Windhorst, R.~A.\ 2003, \apjl, 596, L5 
\bibitem[Conselice et al.(2005)]{con05} Conselice, C.~J., Blackburne, J.A., \& Papovich, C. 2005, ApJ, 620, 564
\bibitem[Conselice(2006)]{con06} Conselice, C.~J.\ 2006, \apj, 638, 686 
\bibitem[Cox et al.(2006)]{cox06} Cox, T.J., et al.\ 2006, \mnras, submitted, astro-ph/0503201
\bibitem[Dale et al.(2001)]{dal01} Dale, D.~A., Helou, G., Contursi, A., Silbermann, N.~A., \& Kolhatkar, S.\ 2001, \apj, 549, 215 
\bibitem[Dasyra et al.(2006)]{das06} Dasyra, K.~M., et al.\ 2006, \apj, 638, 745 
\bibitem[Efron (1981)]{efr81} Efron, B. 1981, Biometrika,68, 589
\bibitem[Efron \& Tibshirani (1986)]{efr86} Efron, B., \&Tibshirani, R. 1986, Stat. Sci., 1,54
\bibitem[Elbaz et al.(2002)]{elb02} Elbaz, D., Cesarsky, C.~J., Chanial, P., Aussel, H., Franceschini, A., Fadda, D., \& Chary, R.~R.\ 2002, \aap, 384, 848 
\bibitem[Fadda et al.(2004)]{fad04} Fadda, D., Jannuzi, B.~T., Ford, A., \& Storrie-Lombardi, L.~J.\ 2004, \aj, 128, 1
\bibitem[Fadda et al.(2006)]{fad06} Fadda, D., et al.\ 2006, \aj, 131, 2859
\bibitem[Fazio et al.(2004)]{faz04} Fazio, G.~G., et al.\ 2004, \apjs, 154, 10 
\bibitem[Kennicutt(1998)]{ken98} Kennicutt, R.~C., Jr.\ 1998, \araa, 36, 189  
\bibitem[Lacy et al.(2004)]{lac04} Lacy, M., et al.\ 2004, \apjs, 154, 166 
\bibitem[Lacy et al.(2005)]{lac05} Lacy, M., et al.\ 2005, \apjs, 161, 41
\bibitem[Lagache et al.(2004)]{lag04} Lagache, G., et al.\ 2004, \apjs, 154, 112
\bibitem[Lavery et al.(2004)]{lav04} Lavery, R.~J., Remijan, A., Charmandaris, V., Hayes, R.~D., \& Ring, A.~A.\ 2004, \apj, 612, 679 
\bibitem[Le F{\`e}vre et al.(2000)]{lef00} Le F{\`e}vre, O., et al.\ 2000, \mnras, 311, 565
\bibitem[Le Floc'h et al.(2005)]{lef05} Le Floc'h, E., et al.\ 2005, \apj, 632, 169
\bibitem[Lilly et al.(1996)]{lil96} Lilly, S.~J., Le Fevre, O., Hammer, F., \& Crampton, D.\ 1996, \apjl, 460, L1  
\bibitem[Lin et al.(2004)]{lin04} Lin, L., et al.\ 2004, \apjl, 617, L9
\bibitem[Lin et al.(2006)]{lin06} Lin, L., et al.\ 2006, \apj, accepted, astro-ph/0607272
\bibitem[Lotz et al.(2006)]{lot06} Lotz, J,~M., Davis, M., Faber, S.~M., Guhathakurta, P., Gwyn, S., Huang, J., Koo, D.~C., Le Floc'h, E. et al.\ 2006, \apj, submitted.
\bibitem[Madau et al.(1998)]{mad98} Madau, P., Pozzetti, L., \& Dickinson, M.\ 1998, \apj, 498, 106 
\bibitem[Melbourne et al.(2005)]{mel05} Melbourne, J., Koo, D.~C., \& Le Floc'h, E.\ 2005, \apjl, 632, L65 
\bibitem[Mihos \& Hernquist(1996)]{mih96} Mihos, J.~C., \& Hernquist, L.\ 1996, \apj, 464, 641
\bibitem[Patton et al.(1997)]{pat97} Patton, D.~R., Pritchet, C.~J., Yee, H.~K.~C., Ellingson, E., \& Carlberg, R.~G.\ 1997, \apj, 475, 29 
\bibitem[Patton et al.(2000)]{pat00} Patton, D.~R., Carlberg, R.~G., Marzke, R.~O., Pritchet, C.~J., da Costa, L.~N., \& Pellegrini, P.~S.\ 2000, \apj, 536, 153 
\bibitem[Patton et al.(2002)]{pat02} Patton, D.~R., et al.\ 2002, \apj, 565, 208 
\bibitem[Patton et al.(2005)]{pat05} Patton, D.~R., Grant, J.~K., Simard, L., Pritchet, C.~J., Carlberg, R.~G., \& Borne, K.~D.\ 2005, \aj, 130, 2043
\bibitem[Rieke et al.(2004)]{rie04} Rieke, G.~H., et al.\ 2004, \apjs, 154, 25 
\bibitem[Sanders et al.(1988)]{san88} Sanders, D.~B., Soifer, B.~T., Elias, J.~H., Madore, B.~F., Matthews, K., Neugebauer, G., \& Scoville, N.~Z.\ 1988, \apj, 325, 74
\bibitem[Schweizer(1982)]{sch82} Schweizer, F.\ 1982, \apj, 252, 455 
\bibitem[Thompson et al.(2001)]{tho01} Thompson, R.~I., Weymann, R.~J., \& Storrie-Lombardi, L.~J.\ 2001, \apj, 546, 694
\bibitem[Shim et al.(2006)]{shi06} Shim, H., Im, M., Pak, S., Choi, P., Fadda, D., Helou, G., \& Storrie-Lombardi, L.\ 2006, \apjs, 164, 435 
\bibitem[Wolf et al.(2005)]{wol05} Wolf, C., et al.\ 2005, \apj, 630, 771 
\bibitem[Xu et al.(2004)]{Xu04} Xu, C.~K., Sun, Y.~C., \& He, X.~T.\ 2004, \apjl, 603, L73 
\bibitem[Yee \& Ellingson(1995)]{yee95} Yee, H.~K.~C., \& Ellingson, E.\ 1995, \apj, 445, 37
\bibitem[Zepf \& Koo(1989)]{zep89} Zepf, S.~E., \& Koo, D.~C.\ 1989, \apj, 337, 34

\end{thebibliography}
\end{document}